\documentclass[aps,prl,twocolumn,showpacs,letterpaper]{revtex4}
\usepackage{graphicx,rotating,epsfig,amsmath,color,dsfont}
\newcommand{\kv}{\mathbf{k}}
\newcommand{\qv}{\mathbf{q}}

\newcommand{\be}{\begin{equation}}
\newcommand{\ee}{\end{equation}}
\newcommand{\bea}{\begin{eqnarray}}
\newcommand{\eea}{\end{eqnarray}}

\newcommand{\bwt}{\begin{widetext}}
\newcommand{\ewt}{\end{widetext}}
\newcommand{\ham}{\mathcal{H}}
\newcommand{\fc}{\mathcal{F}}
\newcommand{\ra}{\rangle}
\newcommand{\la}{\langle}
\newcommand{\bsb}{\begin{subarray}}
\newcommand{\esb}{\end{subarray}}
\newcommand{\largem}{\!\!\!\!\!\!}
\newcommand{\ov}{\bar}
\newcommand{\vecv}[2]{
\left(
 \begin{tabular}{c}
  $#1$ \\
  $#2$
  \end{tabular}
\right)
}

\newcommand{\vech}[2]{
\left(
 \begin{tabular}{c}
  $#1$ \: $#2$
  \end{tabular}
\right)
}

\newcommand{\mat}[4]{
\left(
 \begin{tabular}{cc}
  $#1$ & $#2$ \\
  $#3$ & $#4$
  \end{tabular}
\right)
}

\begin{document}
\title{Survival of the Dirac points in rippled graphene}

\author{Lucian Covaci and Mona Berciu}
\affiliation{
Department of Physics and Astronomy, University of British Columbia, Vancouver,
BC, Canada, V6T~1Z1}

\begin{abstract}
We study the effects of the rippling of the graphene sheet
on the quasiparticle dispersion. This
is achieved using a generalization to the honeycomb lattice of the
Momentum Average approximation, which is accurate for all coupling
strengths and at all energies. We show that even though the position
of the Dirac points may move and the Fermi speed can be
renormalized significantly, quasiparticles with very long lifetimes
survive near the Dirac points even for very strong couplings. 
\end{abstract}

\date{\today}
\maketitle

Graphene~\cite{Novoselov} has been a hot research topic
recently~\cite{review}, primarily due to its Dirac points and the new
paradigm of relativistic-like electron dispersion in their
vicinity. Such dispersion is predicted by almost any hopping model on
the two-dimensional (2D) honeycomb lattice. However, 2D systems should
not have long-range 
order~\cite{Mermin}, and indeed, free-standing graphene sheets are rippled
\cite{ripples}. In epitaxial graphene~\cite{epitaxial}, coupling to substrate
phonons becomes important.  
Both of these can  be modeled as Holstein-like coupling of
electrons to out-of-plane optical phonons~\cite{stauber:2008}. Here we
investigate the effect of such coupling on the
electronic dispersion.

This issue is important because we know from studies of Holstein
polarons on simple cubic-like lattices that even weak-to-moderate
electron-phonon coupling has significant
effects~\cite{Holstein_review}: while a polaron band with infinite
lifetime appears at very low energies, the higher energy spectral
weight broadens considerably. In other words, phonon emission and
absorption leads to very short lifetimes for all higher-energy
states. This raises the possibility that the Dirac points, which are
at high energies above the bottom of the band, may also be ``washed
out'' into an incoherent and featureless background.

We study this problem here, for a single electron.  Of course,
graphene is a half-metal and phonon-mediated electron-electron
interactions may lead to further broadening. We assume that such
effective interactions, like the Coulomb interactions, have little
effect on lifetimes.

Our results show that well-defined Dirac points with long-lived
quasiparticles are preserved even for extremely strong electron-phonon
coupling, where most of the rest of the spectrum is highly
incoherent. Thus, these most interesting features are very robust,
although their energies are shifted somewhat and the slope may be
renormalized substantially. These results justify why one can ignore
the rippling effects and assume a 2D lattice with long-lived
quasiparticles, as has been done so far. This provides a valid
description near the Dirac points, but would fail if the Fermi energy
was anywhere else.

We use a generalization of the Momentum Average (MA)
approximation~\cite{MA1} to calculate the single-electron Green's
function. MA was shown  to be accurate for the entire spectrum (not
just low-energies) for all coupling
strengths and in all dimensions, for such
problems~\cite{MA2,MA3}. This is so because the MA spectral weight obeys
exactly a significant number of sum rules, and is accurate 
for all higher order ones. It can also be
systematically improved~\cite{MA3}. We use here the generalizations of
the MA$^{(0)}$ and MA$^{(1)}$ levels. There is hardly any difference
between their predictions near the Dirac points, showing that
convergence is reached and we need not go to a higher
level. The results shown throughout are from MA$^{(1)}$, whose
spectral weight fulfills exactly the first 8 sum rules.

Consider the  honeycomb lattice, with basis
vectors $\mathbf{a}_{1,2}=3a/2(1,\pm 1/\sqrt{3})$, and the three
nearest neighbors of any site 
defined by 
 $\mathbf{\delta}_1=(a,0)$,
$\mathbf{\delta}_{2,3}=-a/2(1,\sqrt{3})$. In  $\kv$-space, the
Holstein Hamiltonian for an electron coupled to an out-of-plane
optical phonon mode is:
\bea
&&\ham=\sum_\kv (\phi_\kv c_\kv^\dagger d_\kv + h.c.) + \Omega \sum_{\qv}
(b_\qv^{\dagger} b_\qv + B_\qv^\dagger B_\qv) \nonumber \\
&&\nonumber
+\frac{g}{\sqrt{N}}\sum_{\kv,\qv} \left[ c_{\kv-\qv}^\dagger
c_\kv(b_\qv^\dagger+b_{-\qv}) + d_{\kv-\qv}^\dagger 
d_\kv(B_\qv^\dagger+B_{-\qv})\right]\!\!,\:\quad
\label{hamil}
\eea
where $c^\dagger$ and $b^\dagger$($d^\dagger$ and $B^\dagger$) create 
electrons, respectively phonons on the two sub-lattices. The first
term is the kinetic energy of 
 the electron for  nearest-neighbor hopping, with
$
\phi_\kv = -t \sum_{i=1}^3 e^{\imath \kv  \mathbf \delta_i}
$. Generalization to other  models is straightforward. The
second term describes the optical phonons of frequency $\Omega$, for the
two sub-lattices. The Holstein coupling of the electron to phonons on the same
site is described by the last term, $g$ being the coupling
strength. All $\kv$-sums are over the Brillouin zone, defined by the
reciprocal lattice 
vectors $\mathbf{b}_{1,2}=2\pi/3a(1,\pm\sqrt{3})$. We set  $t=1$,
$\hbar=1$. $N\rightarrow \infty$ is the number of 
unit cells. 

Given the bipartite  lattice, the single-electron Green's
function can be defined as a $(2 \times 2)$ matrix:
\bea
\ov{G}(\kv,\omega)\largem &&=\la 0| \vecv{c_\kv}{d_\kv} \hat{G}(\omega)
\vech{c_\kv^\dagger}{d_\kv^\dagger} |0 \ra \nonumber \\
&&= \mat{\la 0|c_\kv \hat{G}(\omega) c_\kv^\dagger | 0 \ra}{\la 0|c_\kv
\hat{G}(\omega) d_\kv^\dagger | 0 \ra}{\la 0|d_\kv \hat{G}(\omega)
  c_\kv^\dagger 
| 0 \ra}{\la 0|d_\kv \hat{G}(\omega) d_\kv^\dagger | 0 \ra},
\eea
where $\hat{G}(\omega)=(\omega+i\delta-\ham)^{-1}$ and  $|0\rangle$ is the
vacuum.

The resolvent for the free electron is
$\hat{G_0}(\omega)=(\omega+i\delta-\ham_0)^{-1}$, where
$\ham_0=\ham|_{g=0} $ . The free electron propagator can be calculated
straightforwardly: 
\be
\nonumber
\ov{G}_0(\kv,\omega)=\mat{G_{0S}(\kv,\omega)}{e^{i\xi(k)}G_{0A}(\kv,
\omega)}{e^{-i\xi(k)}G_{0A}(\kv,\omega)}{G_{0S}(\kv,\omega)}, \ee
where the symmetric and anti-symmetric parts are: 
\be 
\label{G0S}
G_{0S,A}(\kv,\omega)\!=\!\frac{1}{2}\!\!\left(\frac{1}
{\omega+i\delta-|\phi(\kv)|}  
\pm \frac{1}{\omega+i\delta+|\phi(\kv)|} \right) \largem 
\ee while the
phase factor is $e^{i\xi(\kv)}=\phi_\kv/|\phi_\kv|$. As expected in a
bipartite lattice, two symmetric bands arise with energy dispersions
given by $E_{\kv\pm}=\pm |\phi_\kv|$ where $ |\phi_\kv|=\left[ 1+4
\cos^2{\left(\frac{\sqrt{3}}{2}k_ya\right)}+4\cos{\left(\frac{\sqrt{3}}{2}
k_ya\right)\cos{\left(\frac{3}{2}k_xa\right)}}\right]^{1/2} $.
In the first Brillouin zone, the dispersion
vanishes at the two Dirac points $K, K'$ located at $2\pi/3a(1,\pm1/\sqrt{3})$.

We now describe briefly the MA$^{(1)}$ approximation for calculating
$\ov{G}(\kv,\omega)$, emphasizing the differences from the 
derivation of Ref.~\cite{MA3}, due to the two-site basis. Like
there,  first we  generate 
the  equations of motion for this
Green's function, and all the higher order ones it is linked to. This
is achieved by using repeatedly Dyson's equation 
$
\hat{G}(\omega)=\hat{G}_0(\omega)+\hat{G}(\omega) \hat{V} \hat{G}_0(\omega),
$
where $\hat{V}=\ham-\ham_0$ is the electron-phonon interaction. The
first equation is:
\be
\ov{G}(\kv,\omega)=\left[1+ \frac{g}{\sqrt{N}}\sum_{\qv_1}
 \ov{F_1}(\kv,\qv_1,\omega) \right]\ov{G_0}(\kv,\omega),
\label{G}
\ee
where $\ov{F_1}(\kv,\qv_1,\omega)$ is the one-phonon Green's
function:
\be
\nonumber
\ov{F_1}(\kv,\qv_1,\omega)\!=\!\la 0|\! \vecv{c_\kv}{d_\kv}\! \hat{G}(\omega)
\vech{c_{\kv-\qv_1}^\dagger 
b^\dagger_{\qv_1}}{d_{\kv-\qv_1}^\dagger
B^\dagger_{\qv_1}} |0 \ra.
\ee
The equation of motion for $\ov{F_1}$ links it back to $\ov{G}$, but
also to two-phonon Green's functions. The two phonons are
both  on
the same sub-lattice as the electron, or one may be on the other
sub-lattice. The equations for these link them back to
$\ov{F_1}$, but also to a new one-phonon Green's function
$\ov{F_1^*}$, which has the phonon on the different sub-lattice than the
electron. And, of course, to a multitude of three-phonon Green's
functions. 
And so on and so forth. All these
higher-order Green's functions must be proportional to the
$2\times 2$ matrix $\ov{G}(\kv,\omega)$~\cite{MA3}, so it is
convenient to rescale them accordingly, 
{\em e.g.},
$
\ov{f_1}(\kv,\qv_1,\omega)=g\sqrt{N}\ov{G}^{-1}(\kv,\omega)
\ov{F_1}(\kv,\qv_1,\omega), 
$
and similarly for all the other ones  (see below). Combining this
with Eq. (\ref{G}), we find the standard solution
$\ov{G}(\kv,\omega)=\left[\ov{G_0}^{-1}(\kv,\omega) - 
  \ov{\Sigma}(\kv,\omega) \right]^{-1}$, where 
 the self-energy is: 
\be
\ov{\Sigma}(\kv,\omega)=\frac{1}{N}\sum_{\qv_1} \ov{f_1}(\kv,\qv_1,\omega).
\label{self_eng}
\ee
Our task is thus to calculate
the momentum average over the first Brillouin zone of the generalized
Green's function $\ov{f_1}(\kv,\qv_1,\omega)$. As just discussed, this
is the solution of an infinite system of coupled
equations of motion. We cannot solve it exactly, so we proceed to
make simplifications. At the MA$^{(0)}$ level, we replace all free
propagators that appear in all these equations by their momentum
averages over the first Brillouin zone. Within MA$^{(1)}$, we keep the
equations for $\ov{f_1}$ and $\ov{f^\star_1}$ unchanged, and make
the MA approximation only from the second level on,
i.e. for free propagators of energy $\omega-n\Omega$ where $n\ge
2$. Either of these approximations allows us to solve the resulting equations of
motion exactly. We proceed to discuss the MA$^{(1)}$ solution in more detail.

As for the simpler case presented at length in
Ref.~\cite{MA3}, the simplified MA$^{(1)}$ equations of motion can be solved in
terms of total (and partial) momentum averages of
these higher Green's functions, over  all (all minus one) of their
phonons' momenta. After such total (partial) averages, only
contributions from Green's function which have all (all except one) of
their phonons on the same sub-lattice as the electron are
non-vanishing. This is because one can easily check, using Fourier
transforms, that:
$$
\sum_{\qv_1, ..., \qv_n} c^\dagger_{\kv-\qv_1-\ldots-\qv_n} b^\dagger_{\qv_1}
\ldots b^\dagger_{\qv_n} = N^{n-1\over 2} \sum_{i}^{} e^{i\kv \cdot
  \mathbf{R}_i} c^\dagger_i (b^\dagger_i)^n.
$$
If any phonon is on the other sub-lattice, the sum over its
momentum vanishes, since it cannot be at the same site as
the electron. This shows the variational meaning of these
approximations~\cite{Barisic,MA3}: in MA$^{(0)}$, a phonon cloud can
appear at any one site. In MA$^{(1)}$  there can  also be an additional
phonon anywhere else in the system. MA$^{(1)}$ thus describes correctly the
polaron+one-phonon continuum~\cite{MA3}, but this is a very low-energy
feature compared to the Dirac points. Diagramatically, both  sum all the
self-energy diagrams, but each diagram is simplified~\cite{MA3}.

To summarize,
the only Green's functions whose total/partial averages remain finite are
(after rescaling):
\begin{widetext}
\bea
&& \nonumber \ov{f_n}(\qv_1,\ldots,\qv_n)=g^n N^{n\over 2}
\ov{G}^{-1}(\kv,\omega)\la 0 | \vecv{c_\kv}{d_\kv} \hat{G}
\vech{c^\dagger_{\kv-\qv_T} b^\dagger_{\qv_1} \ldots
  b^\dagger_{\qv_n}}{d^\dagger_{\kv-\qv_T} B^\dagger_{\qv_1} \ldots
  B^\dagger_{\qv_n} } |0 \ra \\ 
&& \nonumber \ov{f^\star_n}(\qv_1,\ldots,\qv_n)=g^n N^{n\over 2}
\ov{G}^{-1}(\kv,\omega)\la 0 | \vecv{c_\kv}{d_\kv} \hat{G} \vech{e^{-i
	\xi_{\kv-\qv_1}}d^\dagger_{\kv-\qv_T} b^\dagger_{\qv_1}
  B^\dagger_{\qv_2} \ldots B^\dagger_{\qv_n}}{e^{i \xi_{\kv-\qv_1}}
  c^\dagger_{\kv-\qv_T} B^\dagger_{\qv_1} b^\dagger_{\qv_2}  \ldots
  b^\dagger_{\qv_n} } |0 \ra 
\eea 
\end{widetext}
Here, $\qv_T=\sum_{i=1}^{n} \qv_i$, and for simplicity of notation we
do not write
explicitly the $\kv$, $\omega$ dependence from now on.

We define the total momentum averages $\ov{\fc_{n}}=1/N^{n}\sum_{\qv_1
  \ldots \qv_n} \ov{f}_{n}(\qv_1 \ldots \qv_n)$, and the partial
  momentum averages  $\delta\ov{f}_{n}(\qv_1)=1/N^{n-1}\sum_{\qv_2 \ldots
  \qv_n} \ov{f_{n}}(\qv_1 \ldots \qv_n) - \ov{\fc_{n}}$,
  $\delta\ov{f}_{n}^\star(\qv_1)=1/N^{n-1}\sum_{\qv_2 \ldots \qv_n}
  \ov{f_{n}^\star}(\qv_1 \ldots \qv_n)$. In terms of these, the
  equations of motion are, for any $n\ge 2$:
\bea
&&\ov{\fc}_n=g_{0S,n}\left( n g^2 \ov{\fc}_{n-1} + \ov{\fc}_{n+1}
  \right), \label{av_1}\\ 
&&\largem \largem \delta \ov{f_n}(\qv_1)= g_{0S,n}\left[ (n-1) g^2
  \delta \ov{f}_{n-1}(\qv_1) + \delta \ov{f}_{n+1}(\qv_1) \right]\!,
  \label{av_2}\\ 
 &&\largem \largem \delta\ov{f}_n^\star(\qv_1)=g_{0S,n}\left[ (n-1)
	g^2 \delta\ov{f}_{n-1}^\star(\qv_1) +
	\delta\ov{f}_{n+1}^\star(\qv_1) \right]\!,\label{av_3} 
\eea
where we use the short-hand notation
\bea
g_{0S,n}\equiv g_{0S}(\omega-n\Omega)={1\over N} \sum_{\kv}^{}
G_{0S}(\kv,\omega-n\Omega) \label{MAg} 
\eea where the free
propagator is given in Eq. (\ref{G0S}). For
$N\rightarrow \infty$, this is a simple 2D integral over the Brillouin zone.

Recurrence equations of the type (\ref{av_1})-(\ref{av_3}) have
solutions in terms of the  continued fractions $A_n(\omega)$ \cite{MA3}:
\be
A_n(\omega)=\frac{ng_{0S,n}}{1-g^2g_{0S,n}A_{n+1}(\omega)}.
\ee

In particular:
\bea
&&\ov{\fc_2}=g^2 A_2(\omega) \ov{\fc_{1}}, \label{r1}\\
&&\delta \ov{f}_2(\qv_1)=g^2 A_1(\omega-\Omega) \delta
\ov{f}_1(\qv_1), \label{r2}\\ 
&&\delta\ov{f}_2^\star(\qv_1)=g^2 A_1(\omega-\Omega)
\delta\ov{f}_{1}^\star(\qv_1).\label{r3} 
\eea

These can be combined with the exact equations of motion for the
original (rescaled) 
$\ov{f_1}(\qv_1)$ and $\ov{f^\star_1}(\qv_1)$ one-phonon Green's functions,
which read:

\begin{widetext}
\bea
&&\ov{f_1}(\qv_1)= G_{0S}(\kv-\qv_1,\omega-\Omega) \left[g^2 +
  \frac{1}{N}\sum_{\qv_2}\ov{f_2}(\qv_1,\qv_2)\right] + 
G_{0A}(\kv-\qv_1,\omega-\Omega)
\frac{1}{N}\sum_{\qv_2}\ov{f^\star_2}(\qv_1,\qv_2) \label{f_star_1}\\ 
&&\ov{f^\star_1}(\qv_1)= G_{0A}(\kv-\qv_1,\omega-\Omega) \left[g^2 +
  \frac{1}{N}\sum_{\qv_2}\ov{f_2}(\qv_1,\qv_2)\right] + 
G_{0S}(\kv-\qv_1,\omega-\Omega)
\frac{1}{N}\sum_{\qv_2}\ov{f_2^\star}(\qv_1,\qv_2). 
\label{f_star_2}
\eea
\end{widetext}

From these we can easily derive equations for  $\ov{\fc_{1}}, \delta
\ov{f}_1(\qv_1), 
\delta\ov{f}_{1}^\star(\qv_1)$ which combined with
Eqs. (\ref{r1})-(\ref{r3}) allow us to calculate all these
quantities. In particular, we find
$\ov{\Sigma}_{MA^{(1)}}(\omega)=\ov{\fc_{1}}=
\mathds{1}\cdot\Sigma_{MA^{(1)}}(\omega)$, {\it i.e.} it is diagonal and
momentum independent~\cite{note}, where
\be
\Sigma_{MA^{(1)}}(\omega)=\frac{g^2
  g_{0S}(\tilde{\omega})}{1-g^2g_{0S}(\tilde{\omega}) 
  [A_2(\omega)-A_1(\omega-\Omega)]}
\label{self_eng}
\ee
with $\tilde{\omega}=\omega-\Omega-g^2
  A_1(\omega-\Omega)$.
As a result, we obtain
$\ov{G}(\kv,\omega)=\ov{G}_0(\kv,\omega-\Sigma_{MA^{(1)}}(\omega))$.
This  can now  easily be extended to  lattices
with  even more complex unit cells.

\begin{figure}[b]
\centering
\includegraphics[angle=270,width=0.75\columnwidth]{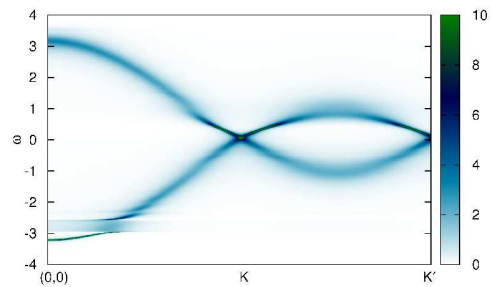}
\caption{(color online) Spectral weight along high-symmetry cuts in
  the Brillouin zone, for $\Omega=0.25t$ and $\lambda=0.5$. } 
\label{fig:1}
\end{figure}

First, we plot in  Fig. ~\ref{fig:1} the spectral weight
$A(\kv,\omega)=-\frac{1}{2\pi} 
Im ({\rm Tr }  \ov{G}(\kv,\omega))$ along high-symmetry cuts in the
BZ. This corresponds to  fairly weak electron-phonon interactions, with
an effective coupling $\lambda = g^2/(3t\Omega)= 0.5$. The 
effective coupling is defined such that the crossover from large to
small polarons is observed for $\lambda \sim 1$, as usual
\cite{Holstein_review}. As expected, the lowest-energy feature is the sharp
polaron band, whose dispersion flattens out just below the
polaron+one-phonon continuum. Because its quasiparticle ({\it qp}) weight
decreases away from $\kv =0$, it is hard to see it in this
region. We have verified that all expected behavior of the
ground-state energy, {\em qp} weight, effective mass, average number
of phonons in the cloud, etc, are indeed similar to those
expected for Holstein polarons. 

Above the polaron band, we see the  polaron+one-phonon continuum,
plus all higher energy features which remain centered around the
corresponding energies
of the free electron. Already there is significant broadening at all
higher energies,  {{\em except near the Dirac points}, which continue to be
very sharp, indicating long lifetimes.

This is further analyzed in Fig.~\ref{fig:2}, where we present the
imaginary part of the self-energy for two different phonon energies,
and for different coupling strengths, in the vicinity of the Dirac
points. Since at the MA$^{(1)}$ level the self-energy is still
momentum independent, the characteristic lifetime is just $\tau \sim
-1/Im(\Sigma)$. The results in Fig.~\ref{fig:2} show that as the
coupling, and thus the electron-phonon scattering, is increased,
lifetimes become shorter and shorter. However, in a relatively narrow
energy interval, the lifetimes remain very long even at extremely
strong couplings which are well into the small polaron regime. In
other words, well defined Dirac points still exist, although their
energy shifts monotonically from $\omega_{K}=0$ at weak coupling to
$\omega_K\sim 2\Omega$ at strong coupling. Note that for very strong
couplings one can also observe the appearance of resonances spaced by
$\Omega$, which signal the Lang-Firsov states, expected when
$\lambda\rightarrow \infty$~\cite{LangFirsov,zoli:2001}.

\begin{figure}[t]
\centering 
\includegraphics[angle=270,width=1.1\columnwidth]{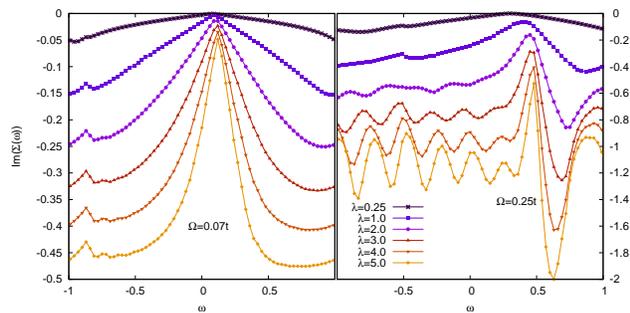}
\caption{(color online) Imaginary part of the self-energy
  for $\Omega=0.07t$ (left) and $\Omega=0.25t$
  (right).} 
\label{fig:2}
\end{figure}

\begin{figure}[b]
\centering 
\includegraphics[angle=270,width=0.75\columnwidth]{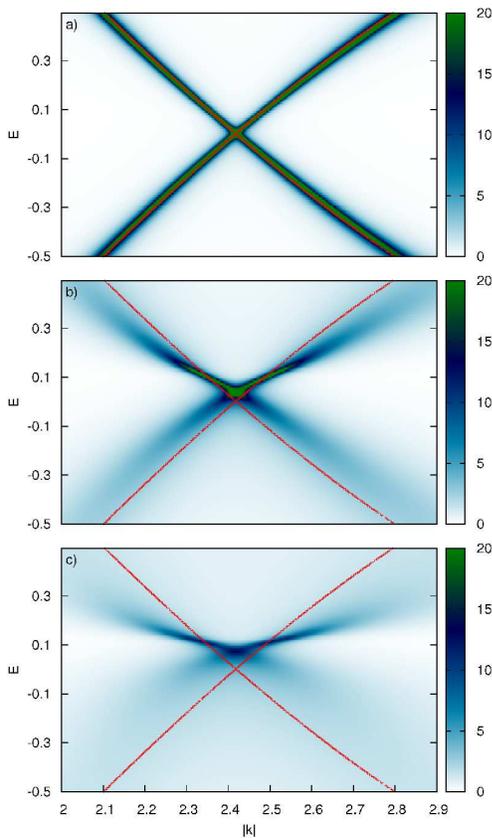}
\caption{(color online) Spectral weight near the Dirac points for a)
  $\lambda=0.25$, b) $\lambda=2.0$ and c) $\lambda=4.0$ and
  $\Omega=0.07t$. The red lines indicate the free-electron dispersion 
  ($\lambda =0$).} 
\label{fig:3}
\end{figure}

Besides the  shift in the energy of the Dirac points, we also
find  significant renormalization (decrease) of the effective
``speed of light'' with increased coupling.  Also, the upper
branch has much longer {\it qp} lifetimes. These points are illustrated
in Fig.~\ref{fig:3}, where we plot the spectral weight for
$\kv$ near the K Dirac point at various couplings, together with the free
electron dispersion (red lines). They also agree with the weak-coupling
results of Ref.~\cite{stauber:2008}.

One may wonder if these results are an artifact due to the
$\kv$-independent self-energy produced by MA$^{(1)}$~\cite{note}. This
cannot be the case, since  MA$^{(1)}$ obeys 
exactly 8 spectral weight 
sum-rules for each value of $\kv$, so significant spectral weight
shifts necessary 
for the disappearance of the Dirac points are simply  impossible. The
explanation for this comes from rewriting Eq. (\ref{MAg}) as $g_{0S}(\omega)
={1\over 2} 
\int_{}^{} d\epsilon g(\epsilon)[1/(\omega- \epsilon +i\eta) +
  1/(\omega+ \epsilon +i\eta)] $, where 
$g(\epsilon)$ is the free-electron density of states (DOS). The largest
contributions come from $\epsilon \sim \pm \omega$, however near the Dirac
points $\omega \sim 0$ and the DOS vanishes. As a result, the self-energy
of Eq. (\ref{self_eng}) remains small near (just above) the Dirac
points, because $g_{0S}(\tilde{\omega})\rightarrow 0$.

We conclude that well-defined Dirac points are a very robust feature
of the graphene, {\em i.e.} rather insensitive to rippling effects
or, for epitaxial graphene, to the nature of
the substrate and the buckling due to mismatching. This is very fortunate,
since it guarantees that the 
interesting physics expected 
because of the Dirac points is not affected by such effects. It also
explains why they can be neglected when studying these quasiparticles,
{\em e.g.} in their interactions with in-plane phonons~\cite{sarma:2007}.

Acknowledgments: This work was supported by the A. P. Sloan Foundation, CIFAR
Nanoelectronics, and NSERC. Discussions with
R. Capaz, A. Castro-Neto and G. A. Sawatzky are gratefully acknowledged.

\bibliographystyle{revtex}

\begin{thebibliography}{99}
\bibitem{note} For Holstein-like coupling, $\kv$-dependent self-energy
  appears from  level 
MA$^{(2)}$  onwards (see Ref.~\cite{MA3}), however this has little
  effect at the 
high-energies of interest here.
\end{thebibliography}

\end{document}